\documentclass{nature-preprint}
\usepackage{tcolorbox,color}
\usepackage{colortbl}
\usepackage{ulem}
\usepackage{amsmath}
\usepackage{amssymb}
\usepackage{graphicx}
\usepackage{upgreek}

\usepackage{caption,wrapfig}
\usepackage{hyperref}
\usepackage{lipsum}
\usepackage[labelfont=bf]{caption}
\usepackage[super,comma,compress]{natbib}

\title{Dynamic Control of Plasmonic Colors by Voltage Actuation MEMS Cantilevers for Optical Display Applications}

\author{Zhengli~Han$^{1,\dagger,*}$, Christian~Frydendahl$^{1,\dagger}$, Noa Mazurski$^{1}$, and Uriel~Levy$^{1,*}$}

\begin{document}

\maketitle 

\begin{affiliations}
\small
\item Department of Applied Physics, The Faculty of Science, The Center for Nanoscience and Nanotechnology, The Hebrew University of Jerusalem, Jerusalem 91904, Israel.
\item[]
\item[$\dagger$] These authors contributed equally to this work
\item[*] han.zhengli@mail.huji.ac.il and ulevy@mail.huji.ac.il
\end{affiliations}


\begin{abstract}
\textbf{Abstract:} Conventional optical displays using ITO (indium tin oxide) and LC (liquid crystal) materials present a lot of challenges in terms of long-term sustainability. We show here how it is possible to generate a cost-effective and CMOS compatible fast and full range electrically controlled RGB color display by combining transmission based plasmonic metasurfaces with MEMS (Microelectromechanical systems) technology, using only two common materials: Aluminum and silicon oxide. White light is filtered into red, green, and blue components by plasmonic metasurfaces made of aluminum nanohole arrays, and the transmission through each color filter is modulated by MEMS miniaturized cantilevers fabricated with aluminum and silicon oxide on top of the color filters. We show that the relative transmission of a color subpixel can be freely modulated from 35\% to 100\%. Our pixels can also operate well above 800\,Hz, enabling future ultrafast displays. Our work provides a road to future circular economic goals by exploiting advances in structural colors and MEMS technologies to innovate optical displays.
\end{abstract}


\section*{Introduction}

Optical displays have become an ubiquitous technology in daily life, being an essential component of smart phones, computers, televisions, etc. These same devices make up roughly half of the 50 million tonnes of e-waste produced every year globally\cite{WEF:2019}. In estimate, less than 20\% of the mass of world's annual e-waste is recorded as being recycled\cite{WEF:2019}. For example, typically only metals are extracted from plastic circuit boards. The remaining 40 million tonnes of waste is mainly unaccounted for, generally ending up in burn pits or land fills in developing countries where many of the toxic elements contained within (such as indium, lead, mercury, etc.) pose serious threats to human and environmental health. Annual e-waste production is predicted to grow to over 120 million tonnes by 2050, with the majority of this increase being from devices reliant on displays\cite{WEF:2019,Bhakar:2015,Lahtela:2019}. One of the main reasons for the low rate of recycling of displays, is the cost associated with the process due to the large variety of materials included in a device\cite{Lahtela:2019}, making 'virgin' materials favorable over recycled materials in terms of cost - despite the fact that it is up to 10 times as energy efficient to recycle e.g. metals over extracting new metals from ore\cite{WEF:2019}. There is thus an urgent need for alternative display technologies that are easier to recycle and are financially feasible to be incorporated into a circular consumer electronics economy\cite{Kneese:1988,Stahel:2016,Kirchherr:2017,Geissdoerfer:2017} - not just to preserve finite resources, but also to lower total energy usage of the manufacturing industry and lower total global carbon emissions\cite{WEF:2019}. In particular, some of the most critical materials to find replacements for are the liquid crystals (LCs) and the indium tin oxide (ITO) which are used for the pixel brightness control in all common digital displays such as LCDs (liquid crystal displays), LEDs (light-emitting diodes displays), and OLEDs (organic light-emitting diodes displays)\cite{Lee:2008,Bhakar:2015,Lahtela:2019}. We show here a new type of optical display technology, based on structural colors and flexible micro mechanical shutters, which could be an ideal candidate to enable the future transition towards a circular electronics economy.

Structural colors generate colors from nanostructures by plasmonic or dielectric optical resonators or gratings\cite{Kristensen:2016,Shao:2018,Lee:2018,Daqiqeh:2020}. While abundant in nature\cite{Kinoshita:2005}, engineered passive structural colors have also been demonstrated to great effect in the last decade\cite{Kristensen:2016,Shao:2018,Daqiqeh:2020}. Both metallic\cite{Kumar:2012,Ellenbogen:2012,Clausen:2014,Goh:2014,Li:2016} and dielectric structures\cite{Zhu:2017,Lee:2018,Yang:2020,Roostaei:2021} have been shown, and generic 'structural color paper' that can be 'painted' using a high-power laser has also been demonstrated\cite{Zijlstra:2009,Chen:2014,Zhu:2016,Zhu:2017,Roberts:2018,Veiko:2021}. Apart from passive structural colors, tunable or 'dynamic' structural colors for display applications are also seeing increasing interest\cite{Shao:2018}. Recent demonstrations include chemical processes to change the local refractive index to shift the resonant color\cite{Duan:2017,Duan:2018}, or using LC as in a conventional LCD monitor to modulate color pixel transmission\cite{Franklin:2017,Shao:2018}. Use of electrostatic forces to align plasmonic resonant metal nanorods to generate colors has also been shown\cite{Greybush:2019}. The disadvantages of these techniques are that either their color switching speed is slow, or it is difficult to individually control pixels, or they do not greatly simplify the composite materials required to manufacture the display - while the number of materials in the display unit is closely related to the end-of-life-cycle management\cite{Lee:2008,Bhakar:2015,Lahtela:2019}.

MEMS (Microelectromechanical systems) technology has been used in the past to generate alternative displays and color filters\cite{Holmstrom:2014,Ma:2015}. MEMS display technologies can generally be categorized into several groups, such as digital micromirror devices\cite{Hornbeck:1997}, laser scanning displays\cite{Tauscher:2010}, interferometric modulator displays\cite{Sampsell:2006}, digital micro-shutters\cite{Hagood:2007}, grating light valves\cite{Bloom:1997}, etc. MEMS generally offers a significant power reduction when compared to most other display technologies, but each actuation technique has its own unique advantages and disadvantages\cite{Holmstrom:2014,Ma:2015}. For example, laser scanning displays offer almost perfect color gamut displays, as the image is generated by scanning three different monochromatic red, green, and blue lasers, but generally suffer from speckle causing imaging artifacts\cite{Holmstrom:2014}.

MEMS combined with structural colors was very recently demonstrated\cite{Holsteen:2019}. The work used SOI (silicon on insulator) wafers to build a silicon grating above a silicon substrate. By controlling the gap in between the grating and substrate, the reflection color was changed among green, yellow and red. This combined approach shows several advantages, such as low power consumption, CMOS-compatible fabrication, etc. However, from the mechanism of the color control, it is difficult to cover the full range of RGB colors, and it is difficult to show black color. Grating based structural colors are also well-known to have issues with viewing angle due to their diffractive nature, and as the device works in reflection mode, it is hard to incorporate a dedicated light source that would allow viewing in low ambient light conditions.

In this paper we bring together plasmonic metasurfaces and MEMS technology to generate transmission type dynamic color control using only two common and easily recyclable materials: Aluminum and silicon oxide (glass). We demonstrate the generation of a full range of RGB color subpixels. The colors are generated by optical transmission based plasmonic metasurfaces consisting of aluminum nanohole arrays, and the relative transmission/brightness of each color subpixel is modulated by a MEMS cantilever made of aluminum and silicon oxide fabricated on top of the metasurface. By using the nanohole array itself as the actuation electrode for the cantilever, application of a bias between the two then causes electrostatic forces to pull the cantilever down and close off light transmission through the plasmonic array - effectively 'shutting it off'. By modulating the actuation voltage, we show how it is directly possible to modulate the overall brightness of each pixel by changing its average transmission in time via fast modulation.

Our method allows for fast pixel switching speeds well in the excess of 800\,Hz (which is significantly faster than currently available displays), and by using the metasurfaces themselves as the actuation electrodes, it is possible to individually address and modulate specific color subpixels - an essential feature for making a RGB display, and eliminates the need for transparent electrodes made of e.g. ITO. Furthermore, as the entire fabrication process is based on lithography and the materials used are generally considered cost-effective and CMOS (Complementary Metal–Oxide–Semiconductor) compatible, it is also expected that our approach could enable significant fabrication cost reductions for mass scale manufacturing of optical displays. 

Our work shows that it is possible to make a fast, full color range display, using just two component materials. This is an important step towards simplifying implementations of digital displays, and a crucial step towards display technologies with simplified end-of-life-cycle reprocessing procedures - essential for achieving future circular economic and e-waste management goals\cite{Kneese:1988,Stahel:2016,Kirchherr:2017,Geissdoerfer:2017}.


\section*{Results}

\begin{figure*}[h]
    \centering
	\includegraphics[width=0.69\linewidth]{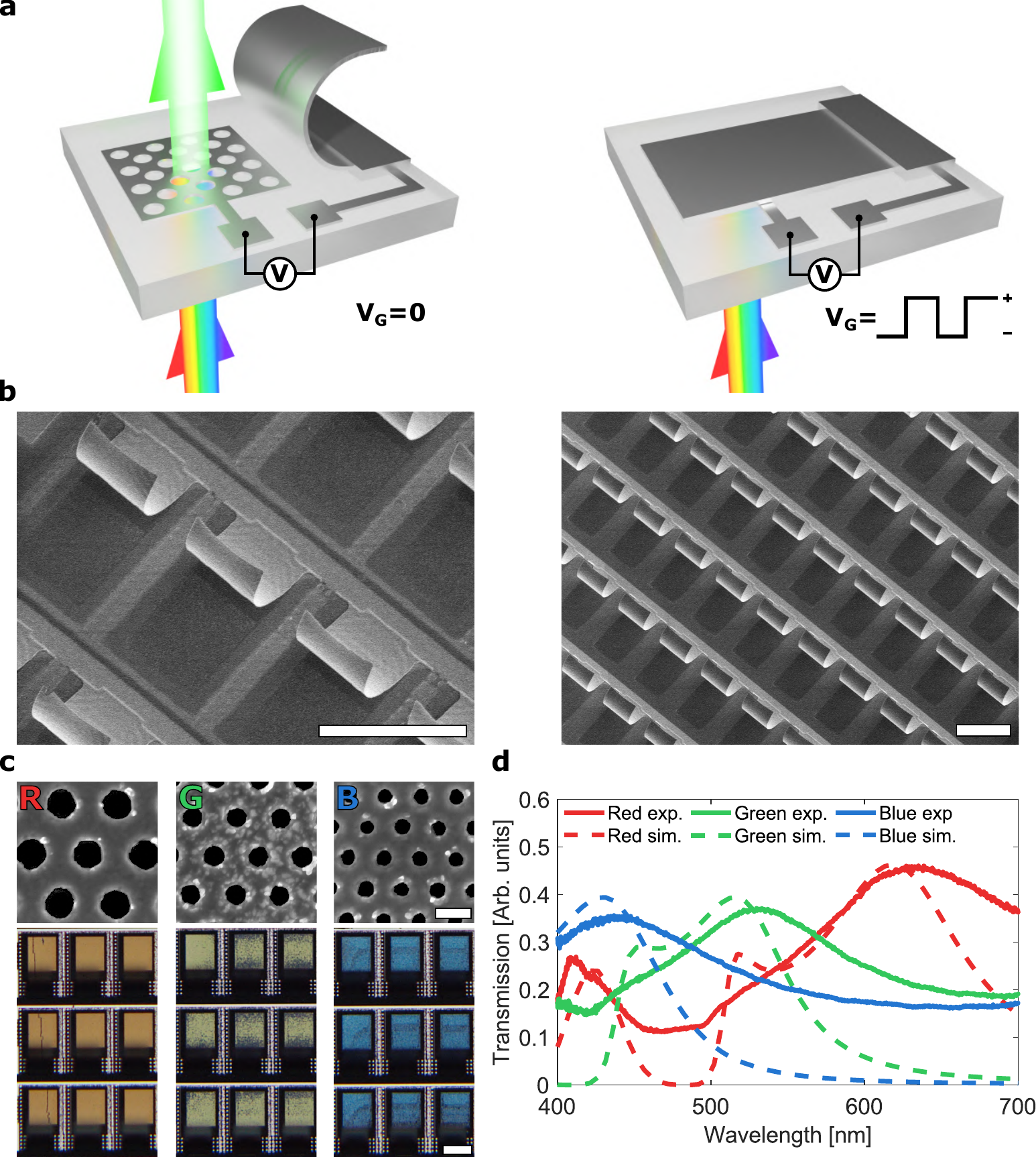}
    \caption{\textbf{Device deign and working principle.} \textbf{a)} Schematic of the device. A plasmonic nanohole array filters white continuum light into one color upon transmission. The array's transmission can then be modulated via a mechanical cantilever which blocks off the array when a voltage is applied between the cantilever and the nanohole array. \textbf{b)} SEM image of the fabricated cantilevers and the nanohole array. Scale bar is 50\,$\upmu$m. \textbf{c)} Top: SEM images of the generated nanohole arrays for the red (R), green (G), and blue (B) color pixel elements. Bottom: White light transmission microscope images for each of the color pixel elements. Scale bars for SEM images are 250\,nm, and for microscope images 30\,$\upmu$m. \textbf{d)} Experimental (solid lines)- and FDTD simulated (dashed lines) transmission spectra of the nanohole arrays.}
    \label{fig:1}
\end{figure*}

\subsection{Device design and operating principle:} 
The general working principle of the device is shown in Fig.~\ref{fig:1}.a. A plasmonic nanohole array filters white continuum illumination into single colors when it is transmitted through the array. Then, the transmission through the array is modulated by a flexible MEMS cantilever, that is actuated by an electrostatic potential applied between the nanohole array and the cantilever. The cantilever motion resembles more the unrolling of a rolled up sheet, rather than a stiff vibrating beam. This allows the metasurfaces to be completely unobstructed by the cantilevers when in their rolled up default position, allowing for high optical transmission. An example of a fabricated device can be seen in the scanning electron microscope (SEM) image in Fig.~\ref{fig:1}.b, and a video of the cantilever motion with low frequency actuation voltage can be seen in the supplementary materials.

To demonstrate the usefulness of our approach towards display applications, we have fabricated three different types of nanohole arrays, one each to match respectively to red, green, and blue visible light\cite{Chen:2010}. By varying the diameter and period of the nanoholes, we are able to change the plasmonic resonance response, and thus control the transmission spectrum at will. The diameters of the nanohles for red, green, and blue are 208\,nm, 154\,nm, and 124\,nm, where the corresponding periods between nanohole center for the three colors are 380\,nm, 312\,nm, and 252\,nm. A set of SEM images and transmission microscope images of the generated arrays can be seen in Fig.~\ref{fig:1}.c, with finite-difference time-domain (FDTD) simulation and experimentally measured transmission spectra available in Fig.~\ref{fig:1}.d. From the spectral measurements we see that the peak transmission efficiencies are $\sim$45\% for the red color filter, and $\sim$35\% for the green and blue color filters. The measured results show noticeable broader spectra at longer wavelength for each curve as compared with the simulation. The details of the spectral measurements and simulations are available in the methods section below.

The overall fabrication of the device is done in two steps: First, the nanohole arrays and a solid electrode are defined and fabricated using electron-beam lithography with an evaporation of a 100\,nm aluminum layer and lift-off. Then, a spacer oxide (to block electrical breakdown when the cantilever is touching the array) of 210\,nm silicon oxide is evaporated on top of the nanoholes. Besides insulating the array from the cantilever, the spacer layer also encapsulates and protects the aluminum nanoholes. In the next step, a sacrificial layer of photoresist is added on top of the array (to later release the cantilevers). Then, using UV photolithography the cantilevers and their electrodes are defined. The cantilevers are then fabricated by sequential sputtering of 100\,nm of silicon oxide, followed by 100\,nm aluminum. Reactive-ion etching (RIE) is then used to etch away the sacrificial layer of photoresist to finally release the cantilevers from the nanohole array and the device substrate. The total lateral size of a nanohole array and its accompanying cantilever is $\sim30\times60$\,$\upmu\text{m}^2$. For more details on the fabrication, see the methods section below.

\subsection{Modulation of color pixel brightness:} 
\begin{figure*}[h]
    \centering
	\includegraphics[width=0.99\linewidth]{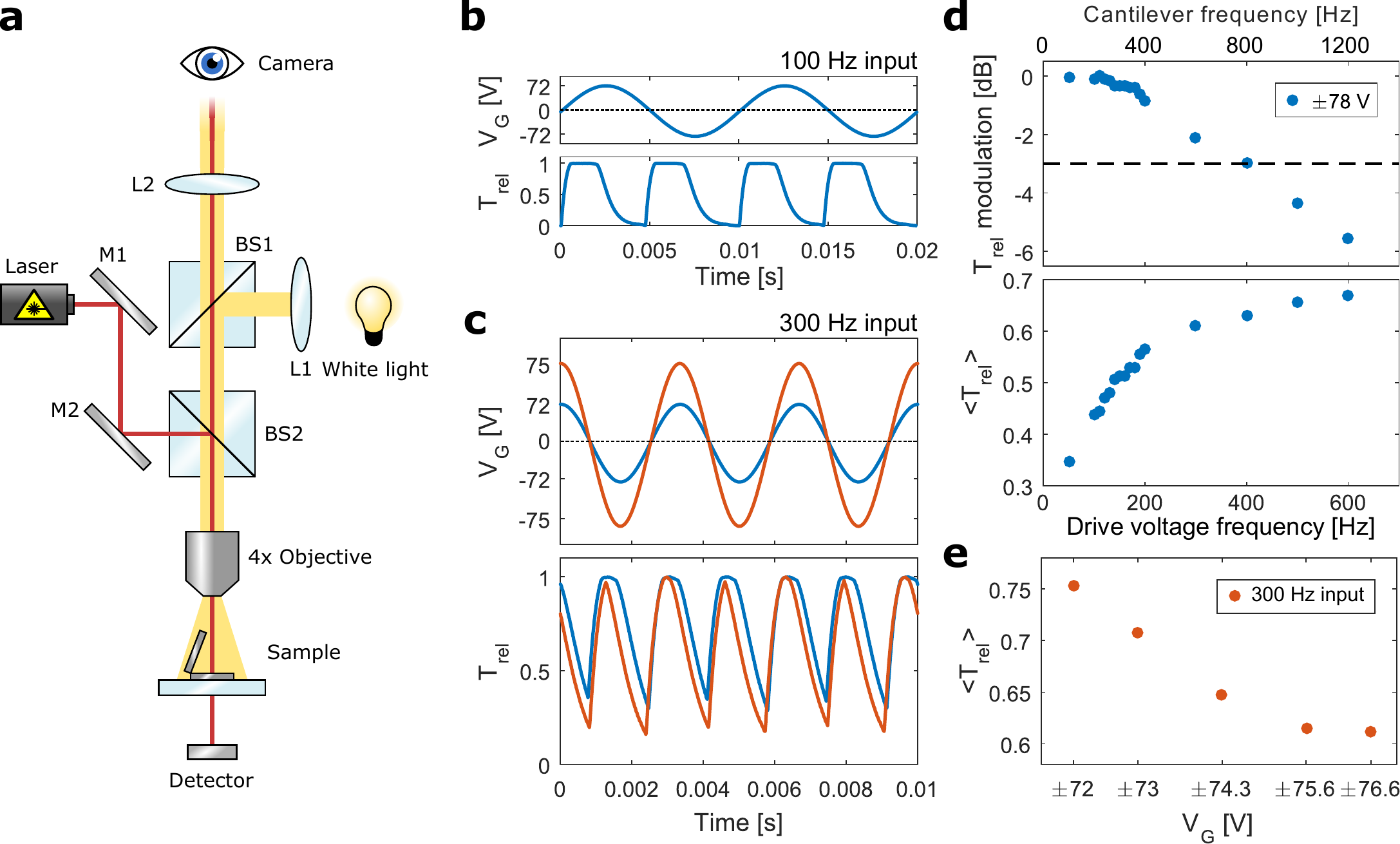}
    \caption{\textbf{Transmission modulation.} \textbf{a)} Experimental setup used for the transmission modulation experiment. A home-built reflection microscope is used to couple laser light through the nanohole array while the cantilevers are modulated by a gate voltage. The resulting transmission modulation is recorded by a silicon photodiode. The imaging part of the microscope (top section) is used to select which cantilever to measure and to align the laser spot to it. \textbf{b - c)} Applied $V_\text{G}$ and the measured relative transmission of a laser passing through the nanohole array, as a function of time, for 100\,Hz and 300\,Hz gate signal, respectively. \textbf{d)} Frequency response of the normalized relative transmission modulation through the array (dotted horizontal line marks -3\,dB) and average relative transmission for different frequencies. \textbf{e)} Average relative transmission of the laser through the array as a function of $V_\text{G}$ peak-to-peak voltages.}
    \label{fig:2}
\end{figure*}

After fabricating the nanohole arrays and cantilevers, we characterize the modulation of light transmission through them using the setup shown in Fig.~\ref{fig:2}.a.

By matching the laser spot size to that of one nanohole array/cantilever we can measure the performance of a single cantilever. The transmission is measured as the photovoltage of a silicon photodiode connected to an oscilloscope. For full details of the transmission modulation measurements, see methods below. It is important to note that the frequency of the light modulation by the cantilevers is doubled relative to the drive voltage frequency, see Fig.~\ref{fig:2}.b. This is because despite the fact that the voltage is applied in alternating patterns of positive and negative voltage, both polarities cause the cantilevers to attract towards the nanohole array. Only when $V_\text{G}$ goes towards zero volt do the cantilevers release back to their natural upward position. This is because the cantilevers are operated by capacitive charging, so when either polarity is applied, there is a corresponding accumulation of positive and negative charges on the nanohole array and the cantilever. This charge accumulation causes the cantilever to actuate down towards the nanohole array, due to electrostatic attraction, regardless of the polarity. The gate voltage is applied in alternating patterns of positive and negative bias to avoid permanent charge accumulation, which would pin the cantilever down permanently if not discharged\cite{Toshiyoshi:2008,han:2014a}. We generally apply a sine-wave pattern for $V_\text{G}$, as the gradual change in polarity allows the cantilevers more time to release and recover. If we apply a square-wave instead, we find that the cantilevers get stuck in the downward position without modulating the light. This is because there is effectively no time at which no bias is applied, and there is no time for the cantilevers to recover to their upward position.

We can write the total force on a cantilever as the sum of a restoring spring force and an attractive electrostatic force. Assuming a parallel plate actuator for simplicity, we get\cite{Toshiyoshi:2008}:
\begin{align}
    F_\text{tot}(x,t) &= F_\text{spring}(x) - F_\text{electrostatic}(x,t),\\
                 &= kx - \frac{1}{2} \epsilon_0 \frac{S}{(g-x)^2} V_\text{G}^2(t),
\end{align}
here $k$ is the spring constant of the cantilever, which is determined by its materials and the geometry, $x$ is its displacement, $\epsilon_0$ is the dielectric constant of vacuum, $S$ is the area of the cantilever, $g$ is the initial gap between the cantilever and the nanohole electrode, and $V_\text{G}$ is the time dependent drive voltage.

We see from Fig.~\ref{fig:2}.c the positive role of increasing the bias. By increasing the amplitude of the drive voltage, $V_\text{G}$, at high frequencies, the 'closed' state of the cantilever becomes deepened (i.e. less light goes through, red versus blue curve in Fig.~\ref{fig:2}.b). While the cantilever is not able to become fully closed (as in Fig.~\ref{fig:2}.b) due to the limited motion speed, a larger electrostatic force from the increased voltage serves to bring it closer to a fully closed position, due to the increased downward acceleration. If we consider the 'well-behaved' cantilever motion at lower frequencies, Fig.~\ref{fig:2}.b, we find also that the rise time of the modulation is $\sim$0.23\,ms while the fall time is $\sim$1.2\,ms. This tells us that the restoring force is generally quite a lot stronger than the electrostatic attractive force, when outside the electrostatic pull-in range of the motion\cite{Toshiyoshi:2008}.

Fig.~\ref{fig:2}.d's top panel shows the peak-to-peak modulation of light in dB through the nanohole array for different drive voltage frequencies. We see from this that the -3\,dB point for the device is at $\sim$400\,Hz input, i.e. $\sim$800\,Hz actual light modulation. For higher frequencies, the peak-to-peak modulation of the light is reduced significantly, as the cantilever does not have time to properly close due to the limited fall-time of the motion.

We can characterize the average transmission of light through the array over time for different frequencies and voltages. For a given measurement of the normalized relative transmission through the sample over time, $T_\text{rel}(t)$, we can calculate the average relative transmission as:
\begin{align}
    \left<T_\text{rel}\right> = \sum_{t=1}^N \frac{T_\text{rel}(t)}{N},
\end{align}
where $N$ is the total amount of equidistant points in time sampled. We can then see that the average transmission through the sample goes up with the drive frequency, bottom panel of Fig.~\ref{fig:2}.d.

Because the speed of the modulation depends on the electrostatic force, we can also modulate the average transmission by changing the peak-to-peak drive voltage, as seen in Fig.~\ref{fig:2}.e.

\begin{figure*}[h]
    \centering
	\includegraphics[width=0.8\linewidth]{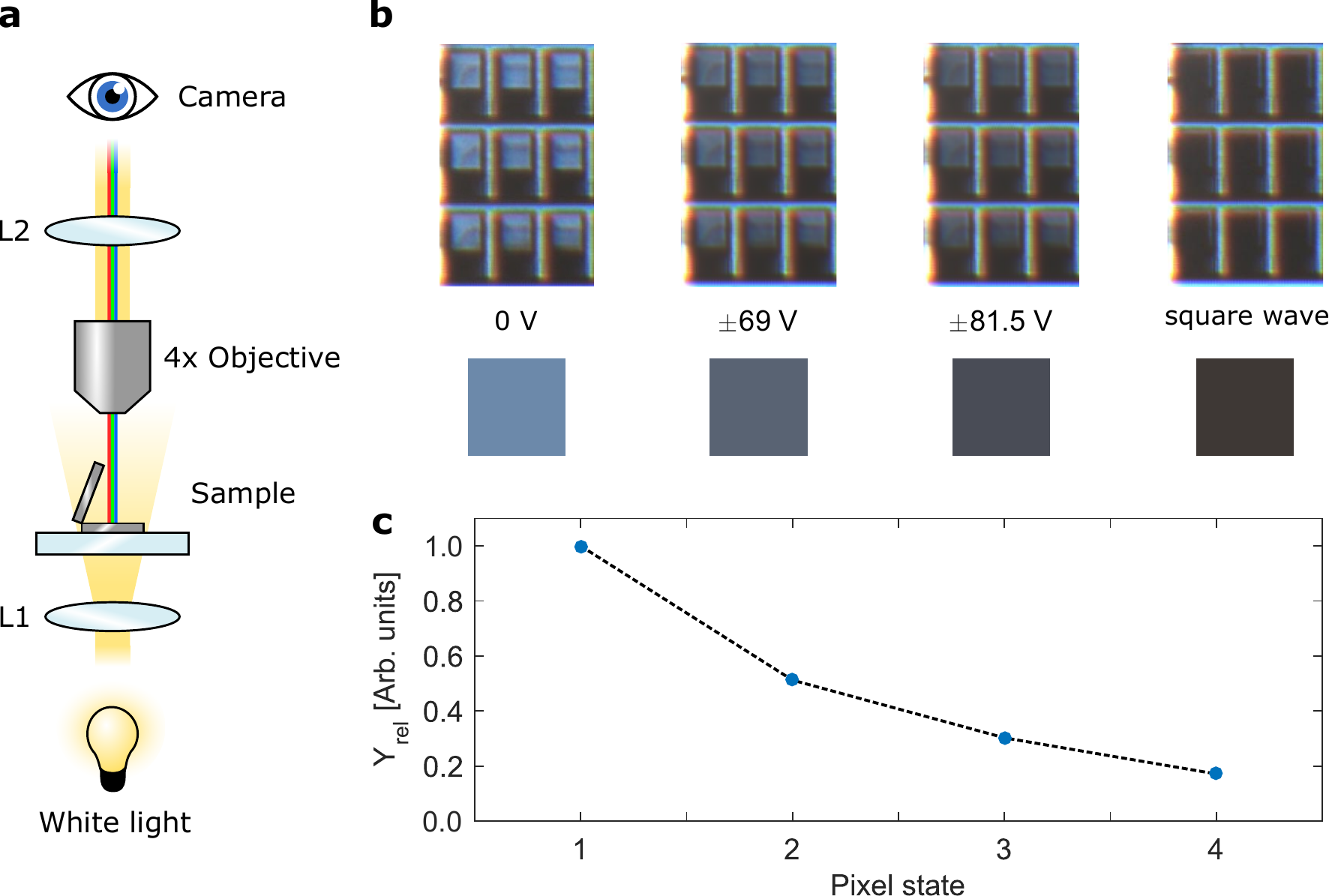}
    \caption{\textbf{Color brightness modulation.} \textbf{a)} Experimental setup used for the brightness measurements. The home-built microscope is reconfigured to a transmission type scheme. \textbf{b)} The resulting brightnesses observed when the cantilevers are modulated at different peak-to-peak voltages with a 100\,Hz driving sine-wave. 100 images were recorded with a framerate of 32\,Hz and averaged together. An average RGB color is also displayed from within the color pixel area. \textbf{c)} Relative luminance value from the averaged RGB values when converted to xyY, normalized to the unmodulated pixel.}
    \label{fig:3}
\end{figure*}

If the motion of the cantilevers is generally faster than the flicker fusion threshold\cite{Landis:1954} of an observer, i.e. the point at which periodic motion/light modulation can no longer be discretely observed, then the modulated response will instead look like the averaged transmission signal. For optical displays, the flicker fusion threshold is generally regarded as being $\sim$60-100\,Hz (different viewing conditions can affect the threshold and it is different for each individual)\cite{Landis:1954}. Motivated by this concept, we decide to test if indeed our device is able to modulate the overall brightness of the transmitted color, as would be necessary for optical display applications, by applying a periodic bias with different gate voltage.

The results can be seen in Fig.~\ref{fig:3}. Using the transmission imaging setup as outlined in Fig.~\ref{fig:3}.a, we can observe the different brightnesses of the color arrays simply by varying the amplitude of drive voltage, while maintaining a 100\,Hz input sine-wave, as seen in Fig.~\ref{fig:4}.b and c.

Because the frame rate of our camera (32\,Hz) is slower than the modulation speed of the cantilevers, we record a series of images (100) and average them together to approximate how a human eye would see the blurred motion of the cantilever. The achieved colors in Fig.~\ref{fig:3}.b are thus only a rough approximation of how a human observer would see the dimming of the pixels.

\subsection{Color mixing:}

\begin{figure*}[h]
    \centering
	\includegraphics[width=0.6\linewidth]{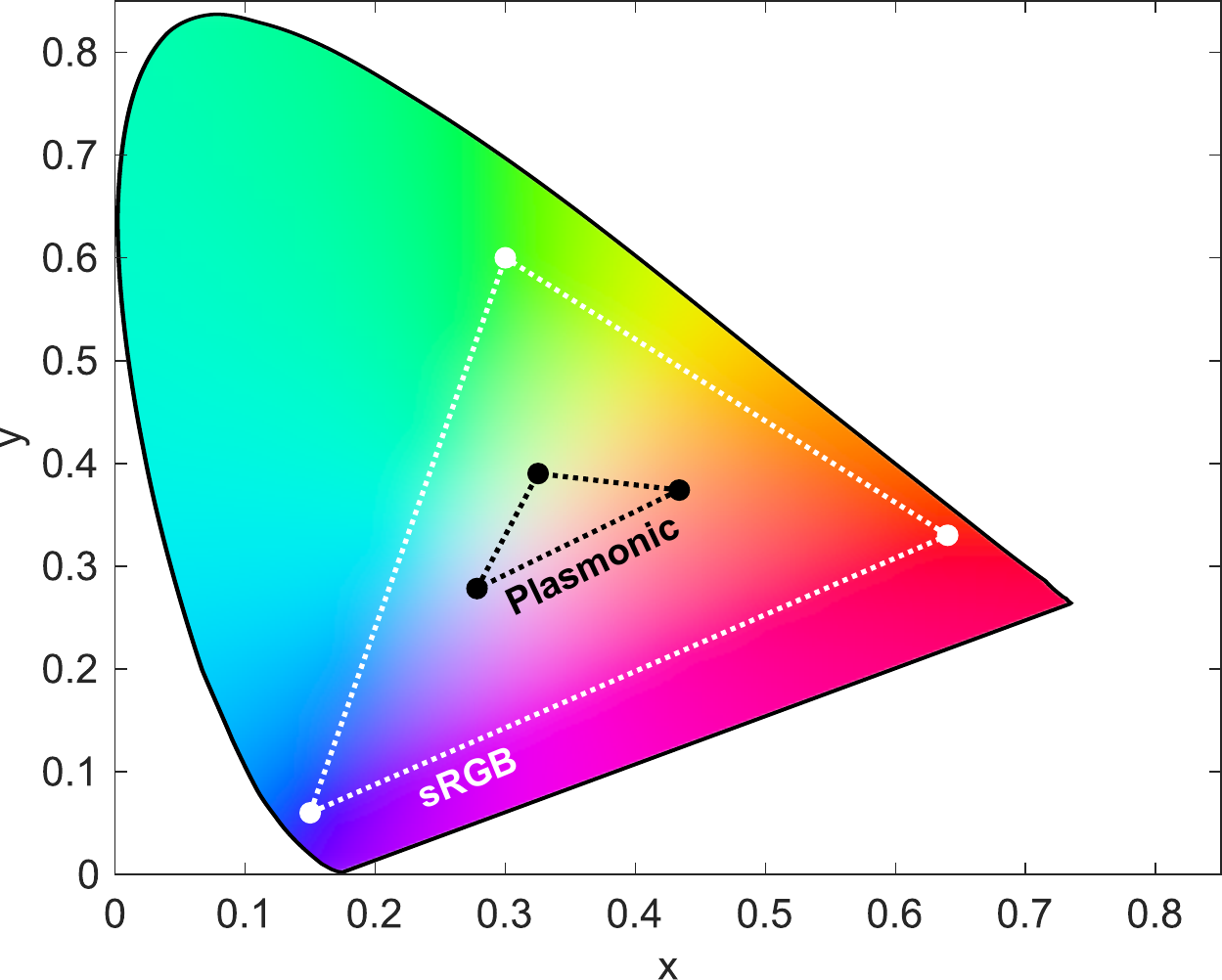}
    \caption{\textbf{Chromaticity and color space} CIE xyY chromaticity map calculated from transmission spectra of white light transmitted through the red, green, and blue color pixels. The range of sRGB colors are shown for comparison.}
    \label{fig:4}
\end{figure*}

Finally, we have also characterized the chromaticity of the red, green, and blue color arrays, the result can be seen in Fig.~\ref{fig:4}.a. The recorded transmission spectra were converted into the International Commission on Illumination (CIE) 1931 color space xyY coordinates (see methods below).

If the colors that make up two points in the CIE xyY chromaticity diagram can be freely mixed, then any colors on a straight line between those two points can also be achieved. Thus, inside a triangle of points, if all three colors corresponding to the points can be mixed freely, then all the colors inside the triangle can be achieved. For comparison, we have shown here also the chromaticity points of the sRGB scale, which is the standard chromaticity scale for most commercial optical displays.

\section*{Discussion}

The capabilities of our 'proof of concept' device shown here could be further improved in several ways. First, the cantilevers are being actuated in air, and drag from air resistance is thus limiting the speed of their motion. Encapsulating the device and operating it in a standard low vacuum of $\sim$10$^{-3}$\,torr could greatly increase the modulation frequencies that could be achieved.

Secondly, the mechanical properties of the cantilevers could be further optimized. It is very possible that by changing the dimensions or the thicknesses of the SiO$_2$ and Al layers, that even higher frequencies of modulation could be achieved, and that the average transmission could be better controlled as well. In the same fashion, adding a high frequency modulation to the drive voltage could further stabilize the cantilever movement, or prevent potential high frequent damping\cite{han:2014a,han:2014b,Toshiyoshi:2008}. The magnitudes of the driving voltages used here (roughly $\pm70$\,V) is also something that could be greatly reduced through optimization of the cantilever design, especially by reducing the cantilever size both in width and length, such that after device fabrication the gap between the cantilever and the bottom electrode will be reduced, which will dramatically lower the pull-in voltage. Another method to reduce the driving voltage is to reduce the spring constant by increasing the cantilever aspect ratio so less total electrostatic force is needed to pull down the cantilever. It should be noted that the spring constant should be kept above a certain level to avoid cantilever stiction, which is a common issue in low spring constant MEMS devices\cite{han:2014a,han:2014b,Toshiyoshi:2008}. 

As for the plasmonic nanohole arrays, further optical design with the aim to narrow the spectral transmission would greatly expand the achievable color space. For example, our red nanohole array contains a great deal of 'blue' light in its transmission spectrum (see Fig.~\ref{fig:1}.d), and in general the arrays allow for a low amount of 'white light' transmission. The color filtering mechanism shown here relies purely on plasmonic resonances. Alternatively, making a combination of diffractive and plasmonic resonances could potentially achieve ultra narrow color filters, as has been shown recently in the short wave infrared regime\cite{Bin:2021}. The on/off extinction ratio of our method is also expected to be very high, as the aluminum cantilever can essentially shut off optical transmission completely, allowing for high contrast ratios in display applications.

In terms of scalability, our approach is also expected to enable cost-effective large scale fabrication of color displays. For one, the entire device only relies on three lithography steps for its fabrication, and two, all of the used materials are CMOS-compatible and considered to be low-cost. While we have utilized e-beam lithography for our particular device, it is possible in industrial quality extreme UV (EUV) lithography to generate structures in the range of 5\,nm\cite{Yeap:2019}, which is well below any feature size in our device.

As our device only consists of aluminum and glass, we believe it would be relatively easy to recycle via a combination of mechanical and heating processes, and/or chemical processes. Essentially the color/display element of a display could be crushed to a fine powder in a mechanical press. Then, as the melting temperatures of glass and aluminum are significantly different ($T_\text{melt,Al}=660$\,$^\circ$C vs. $T_{\text{melt,SiO}_2}=1723$\,$^\circ$C), the aluminum and glass components could be melted out of the powder at different temperatures in a furnace. Alternatively, the powder could be suspended in an acid  which would selectively dissolve the aluminum to form an aluminum salt dissolved in the liquid. The resulting liquid could then be strained of the remaining glass particles, and the aluminum could be extracted either by electro-chemical processes or chemical reduction\cite{Lahtela:2019}.

\section*{Conclusion}

To conclude, we have shown that it is possible to generate a full range of red, green, and blue color pixels using plasmonic metasurface in the form of a nanohole arrays and demonstrate the modulation of the average transmission of such an array by a micromechanical cantilever which is actuated by applying an electrostatic potential between the nanohole array and the cantilever.

We find that the -3\,dB point for the transmission modulation to be $\sim$800\,Hz, which is well above the speeds of conventional display technology (60-124\,Hz), and even more importantly, well above the human flicker fusion threshold (60-100\,Hz). Taking advantage of this property, we have shown that the average transmission through an array can be controlled by changing the magnitude of the applied voltage or by changing the drive voltage frequency.

Making a pixel that consists of three subpixels each of a red, green, and blue nanohole array can thus allow for color mixing between the red, green, and blue subpixels, by individually controlling their average transmission, making their brightness dim to a human observer.

Finally, from transmission spectra of each of our nanohole arrays, we have extracted the chromaticity and shown the color space that could be achieved in such a plasmonic color display.

Our results highlight how plasmonic structural colors and MEMS technology can be combined in order to generate a color display, based on only two materials that can be fully recyclable.  Given the immense prevalence and importance of displays in modern society, such reduction of component materials is predicted to become critically important in the future if circular economic goals of easily, and economically viably, recyclable consumer goods are to be achieved.


\section*{Methods}
\subsection{FDTD simulation:} FDTD simulations were carried out using the commercial software Lumerical FDTD (Ansys). The nanohole arrays were simulated using periodic boundary conditions for the $x$- and $y$-directions and perfectly matched layers (PMLs) were used in the $z$-direction. The smallest mesh element size was 2\,nm. The dielectric functions of the aluminum used the values from Palik \cite{Palik:1998}, and the glass was kept as a lossless dielectric with refractive index 1.45.

\subsection{Sample fabrication:} The device fabrication includes two lithography techniques, electron beam (EB) lithography and photo lithography. We use the EB lithography to make the nanohole arrays for the color filters, and the photo lithography for the MEMS cantilevers. First, PMMA was coated on a glass substrate, followed by the EB lithography with a ELS-G100 (Elionix) system. An acceleration voltage of 100\,kV and a beam current of 1\,nA were used. Then 100\,nm thick Aluminum (Al) was deposited with electron-beam evaporation on the wafer. We used lift-off to pattern the nanohole arrays and the bottom electrodes for the cantilevers. A thin layer oxide of 210 nm was deposited with sputtering on the wafer to tune the color filter of nanohole arrays to the desired color. The oxide layer also works as an insulation layer for the electrodes when actuating the cantilevers. Next, a photoresist sacrificial layer was patterned on the wafer with photo lithography using a MA6 mask aligner (SÜSS MicroTec). A 100\,nm oxide and 100\,nm Al were then deposited with sputtering on the wafer. The cantilever shapes were patterned by photo lithography, followed by Al wet etching and oxide wet etching. The cantilevers are released by O$_2$ plasma with a RIE machine.

\subsection{Transmission spectroscopy:} We characterize the optical transmission of the plasmonic metasurfaces by illuminating them with a white light source (tungsten-halogen lamp) through a microscope condenser lens. The transmitted light is collected with a microscope objective (Nikon, 50×, NA 0.45) and directed into an Ocean Optics Flame spectrometer. An aperture is placed in the image plane to spatially select only the metasurface region (around diameter 25\,$\upmu$m circular area) to enter the spectrometer.

\subsection{Light modulation measurements:} The cantilevers' modulation of light was measured using a red laser pointer. Following the setup in Fig.~\ref{fig:2}.a, the laser light was coupled from above in a home-built reflection microscope, where it was aligned to a single cantilever at a time. Tungsten probes were mechanically pressed against the aluminium electrodes on the top of the device using XYZ 500MIM 3D stages (Quater Research \& Development), and a bias voltage as applied between them using a SDG 2082X (SIGLENT) function generator. The voltage output of the signal generator was amplified using a model 2350 high voltage amplifier (TEGAM). The transmitted laser light through the nanohole array  was then measured using a silicon photodiode (DET100A/M, Thorlabs). The voltage output of the photodiode and the applied bias were both monitored on an InfiniiVision DSOX2004A (Keysight) oscilloscope. The measured transmission signals were converted to $T_\text{rel}$ signals of values between 0 and 1, by first subtracting the off-state voltage from the lowest frequency measurement, and then normalizing the signals to the highest value of the lowest frequency measurement.

\subsection{Color mixing and color gamut:} CIE 1931 xyY chromaticity coordinates were calculated from the recorded transmission spectra. Using the CIE 1931 color matching functions, $\Bar{x}$, $\Bar{y}$, and $\Bar{z}$\cite{Guild:1931} the tristimulus values, $X$, $Y$, and $Z$ were then calculated as\cite{Malacara:2002}:
\begin{align*}
    &X = \frac{1}{N}\int_{0}^{\infty} I(\lambda) T(\lambda) \Bar{x}(\lambda) d\lambda, &Y = \frac{1}{N}\int_{0}^{\infty} I(\lambda) T(\lambda) \Bar{y}(\lambda) d\lambda,& &\text{and}& &Z = \frac{1}{N}\int_{0}^{\infty} I(\lambda) T(\lambda) \Bar{z}(\lambda) d\lambda,
\end{align*}
where $I$ is the spectral power distribution of the illuminating light source, $T$ is the normalized transmission spectrum of a nanohole array color filter, $\lambda$ is the wavelength, and $N$ is given as:
\begin{align*}
    N = \int_{0}^{\infty} I(\lambda) \Bar{y}(\lambda) d\lambda.
\end{align*}
The 'brightness' (luminance) and 'color' (chromaticity) of the light transmitted through the nanohole array can then be described by the CIE xyY coordinates. The tristimulus value $Y$ from above is the luminance per definition, and the $x$ and $y$ chromaticity coordinates are given by normalization as:
\begin{align*}
    &x = \frac{X}{X+Y+Z}, &\text{and}& &y= \frac{Y}{X+Y+Z}.
\end{align*}
The final coordinate $z = Z/(X+Y+Z) = 1 - x - y$ is not independent and is generally disregarded.

\vspace{12pt}

\bibliography{main}

\subsection{Acknowledgments:} The authors wish to thank Mr. Maurice Saidian and Dr. Shimon Eliav of the The Hebrew University Center for Nanoscience and Nanotechnology for assistance with the device fabrication.

\textbf{Funding:} We acknowledge funding from the Israeli Ministry of Science and Technology. Z.H. is supported by the PBC Fellowship Program. 

\subsection{Author contributions:} The idea for the project was devised by Z.H., C.F. and U.L. Optical design and FDTD simulations were done by Z.H. Sample fabrication was done by Z.H. and N.M. SEM images were recorded by N.M. Optical characterization was done by C.F. and Z.H. Data analysis was done by C.F. The work was supervised by U.L. All authors contributed to the writing of the manuscript and interpretation of the results.

\subsection{Conflicts of interests:} 
The authors declare no competing financial interests.

\subsection{Data availability:} Optical spectra and other raw data are available upon request.

\subsection{Correspondence:} General correspondence should be addressed to: han.zhengli@mail.huji.ac.il and\\ ulevy@mail.huji.ac.il


\end{document}